\documentclass[prd,amsmath,amssymb,superscriptaddress,preprintnumbers,twocolumn,nofootinbib,10pt]{revtex4-1}
\pdfoutput=1
\usepackage{graphicx}
\usepackage{dcolumn}
\usepackage{bm}
\usepackage{amssymb}
\usepackage{latexsym}
\usepackage{booktabs}
\usepackage{amsmath}
\usepackage{multirow}
\usepackage{url}
\usepackage{footnote}
\usepackage{float}
\usepackage{threeparttable}
\usepackage[colorlinks=true, linkcolor=blue, citecolor=blue]{hyperref}
\usepackage[bottom]{footmisc}

\usepackage[normalem]{ulem}
\usepackage{color}
\usepackage{array}
\usepackage{enumerate}
\usepackage{adjustbox}

\usepackage{makecell}
\usepackage{diagbox}
\usepackage{epstopdf}
\usepackage{epsfig}
\usepackage{longtable}
\usepackage{supertabular}
\usepackage{algorithm}
\usepackage{pifont}
\usepackage{algorithmic}
\usepackage{changepage}
\usepackage{setspace}

\begin{document}

\title{Revisiting holographic dark energy after DESI 2024}

\author{Tian-Nuo Li}
\affiliation{Liaoning Key Laboratory of Cosmology and Astrophysics, College of Sciences, Northeastern University, Shenyang 110819, China}

\author{Yun-He Li}
\affiliation{Liaoning Key Laboratory of Cosmology and Astrophysics, College of Sciences, Northeastern University, Shenyang 110819, China}

\author{Guo-Hong Du}
\affiliation{Liaoning Key Laboratory of Cosmology and Astrophysics, College of Sciences, Northeastern University, Shenyang 110819, China}

\author{Peng-Ju Wu}
\affiliation{School of Physics, Ningxia University, Yinchuan 750021, China}

\author{Lu Feng}
\affiliation{College of Physical Science and Technology, Shenyang Normal University, Shenyang 110034, China}
\affiliation{Liaoning Key Laboratory of Cosmology and Astrophysics, College of Sciences, Northeastern University, Shenyang 110819, China}

\author{Jing-Fei Zhang}\thanks{Corresponding author}\email{ jfzhang@mail.neu.edu.cn}
\affiliation{Liaoning Key Laboratory of Cosmology and Astrophysics, College of Sciences, Northeastern University, Shenyang 110819, China}

\author{Xin Zhang}\thanks{Corresponding author}\email{zhangxin@mail.neu.edu.cn}
\affiliation{Liaoning Key Laboratory of Cosmology and Astrophysics, College of Sciences, Northeastern University, Shenyang 110819, China}
\affiliation{MOE Key Laboratory of Data Analytics and Optimization for Smart Industry, Northeastern University, Shenyang 110819, China}
\affiliation{National Frontiers Science Center for Industrial Intelligence and Systems Optimization, Northeastern University, Shenyang 110819, China}

\begin{abstract}

{New insights from the Dark Energy Spectroscopic Instrument (DESI) 2024 baryon acoustic oscillations (BAO) data, in conjunction with cosmic microwave background (CMB) and Type Ia supernova (SN) data, suggest that dark energy may not be a cosmological constant. In this work, we investigate the cosmological implications of holographic dark energy (HDE) and interacting holographic dark energy (IHDE) models, utilizing CMB, DESI BAO, and SN data. By considering the combined DESI BAO and SN data, we determine that in the IHDE model, the parameter $c > 1$ and the dark-energy equation of state $w$ does not cross $-1$ at the $1\sigma$ confidence level, whereas in the HDE model, it marginally falls below this threshold. Upon incorporating CMB data, we observe that in the HDE model, the parameter $c < 1$ and $w$ crosses $-1$ at a level beyond $10\sigma$. Conversely, for the IHDE model, the likelihood of $w$ crossing $-1$ is considerably diminished, implying that the introduction of interaction within the HDE model could potentially resolve or mitigate the cosmic big rip conundrum. Furthermore, our analysis reveals that the HDE and IHDE models are statistically as viable as the $\Lambda$CDM model when assessing Bayesian evidence with DESI BAO data combined with SN data. However, when CMB data are added, the HDE and IHDE models are significantly less favored compared to the $\Lambda$CDM model. Our findings advocate for further exploration of the HDE and IHDE models using forthcoming, more precise late-universe observations.}

\end{abstract}
\maketitle

\section{Introduction}

Since the initial discovery of the universe's accelerating expansion \cite{SupernovaSearchTeam:1998fmf,SupernovaCosmologyProject:1998vns}, the nature of dark energy has persisted as an enduring enigma in contemporary cosmology and fundamental physics. Over the past two decades, a plethora of theoretical and phenomenological models have been crafted in response. Notably, the standard $\Lambda$ cold dark matter ($\Lambda$CDM) cosmological model, with six basic parameters, has garnered significant acclaim for its exceptional fit to the majority of cosmological observations.

However, as the precision of cosmological parameter measurements has increased, some perplexing discrepancies have emerged. Particularly, the Hubble constant $H_0$ inferred from the $\Lambda$CDM model's best fit to the Planck cosmic microwave background (CMB) data \cite{Planck:2018vyg} is in significant tension, exceeding $5\sigma$, with the direct measurements by the SH0ES team \cite{Riess:2021jrx} utilizing the distance ladder method. Lately, the Hubble tension has been a hot topic in current cosmology (see, e.g., Refs.~\cite{Li:2013dha,Zhang:2014dxk,Bernal:2016gxb,Feng:2017nss,Zhao:2017urm,Guo:2018ans,Verde:2019ivm,Riess:2019qba,Vagnozzi:2019ezj,Gao:2021xnk,DiValentino:2021izs,Cai:2021wgv,Schoneberg:2021qvd,Kamionkowski:2022pkx,Vagnozzi:2023nrq,Lynch:2024hzh,Huang:2024erq,Huang:2024gfw,Gong:2024yne,Freedman:2024eph}; for relevant forecast analyses, see also, e.g., Refs.~\cite{Zhao:2010sz,Cai:2017aea,Du:2018tia,Zhang:2019ylr,Zhang:2019loq,Chen:2020dyt,Chen:2020zoq,Bian:2021ini,Wang:2021srv,Qi:2022sxm,Zhao:2022yiv,Jin:2022qnj,James:2022dcx,Wu:2021jyk,Song:2022siz,Muttoni:2023prw,Zhang:2023gye,Jin:2023sfc,Jin:2023tou,Pierra:2023deu}). On another front, the cosmological constant 
$\Lambda$ in the $\Lambda$CDM model, which is equivalent to the vacuum energy density, also faces severe theoretical challenges, namely, the ``fine-tuning" and ``cosmic coincidence" problems \cite{Sahni:1999gb,Bean:2005ru}.

In light of the first-year data release from the Dark Energy Spectroscopic Instrument (DESI) collaboration, the $\Lambda$CDM model is encountering heightened scrutiny. The combination of DESI baryon acoustic oscillations (BAO) data with CMB anisotropy measurements from Planck and the Atacama Cosmology Telescope (ACT), along with Type Ia supernova (SN) data from the PantheonPlus, Union3, and DESY5 datasets, has revealed a statistically significant inclination towards a dynamical dark energy scenario, specifically for the $w_0w_a$CDM model, with confidence levels of $2.5\sigma$, $3.5\sigma$, and $3.9\sigma$ respectively. The dark-energy equation of state (EoS) in the $w_0w_a$CDM model is articulated by
\begin{equation}\label{eq1}
w(a)=w_0+w_a(1-a),
\end{equation}
where $a$ represents the cosmological scale factor, and $w_0$ and $w_a$ are constants. The DESI collaboration has reported $w_0 = -0.727 \pm 0.067$ and $w_a = -1.05^{+0.31}_{-0.27}$ based on CMB+DESI+DESY5 datasets, signifying a pronounced preference for $w_0 > -1$ and a substantial $w_a < 0$. These notable deviations from the $\Lambda$CDM model, as indicated by DESI, have ignited extensive debates on the nature of dark energy. Numerous studies have endeavored to constrain cosmological parameters utilizing the novel DESI BAO data \cite{DESI:2024aqx,Wang:2024pui,Escamilla-Rivera:2024sae,DiValentino:2024xsv,Yang:2024kdo,Wang:2024dka,Gomez-Valent:2024tdb,Allali:2024anb,Qu:2024lpx,Wang:2024rjd,Colgain:2024xqj,Cortes:2024lgw,Wang:2024hks,Jiang:2024viw,Giare:2024syw,Giare:2024gpk,Du:2024pai,Toda:2024ncp,Pang:2024qyh,Reboucas:2024smm,Sabogal:2024yha,Escamilla:2024ahl,Li:2024qso,Specogna:2024euz,Chan-GyungPark:2024brx,Park:2024jns,Alestas:2024eic,Wang:2024tjd,Giare:2024ocw,Wu:2024faw,Ye:2024ywg,Tyagi:2024cqp,Li:2025owk,Du:2025iow,Feng:2025mlo}.

While parameterization approaches such as the $w_0w_a$CDM model are valuable for data-driven description and serve as stringent consistency tests against the null hypothesis of the $\Lambda$CDM model, they remain empirical and lack a fundamental physical underpinning. Thus, it is imperative to delve into dark energy models that are anchored in solid theoretical frameworks and to evaluate whether the DESI BAO data can accommodate these models. More fundamentally, it is widely posited that the conundrums surrounding dark energy are inextricably tied to the realm of quantum gravity. Consequently, probing the nature of dark energy through the perspective of quantum gravity is of paramount interest. In the absence of a comprehensive quantum gravity theory, we are compelled to lean on the holographic principle within quantum gravity to construct an effective theory of dark energy.

The holographic dark energy (HDE) model, rooted in the holographic principle of quantum gravity, has gained significant traction in the quest to unravel the enigma of dark energy. Within the framework of effective quantum field theory, \citet{Cohen:1998zx} articulated that, when gravity is taken into account, the total energy of a system with scale $L$ should not surpass the mass of a black hole of equivalent size, that is, $L^{3}\rho_{\rm de}\lesssim LM_{\rm pl}^{2}$. This energy bound gives rise to the density of HDE,
\begin{equation}\label{2.1}
   \rho_{\rm de}=3c^{2} M_{\rm pl}^{2}L^{-2},
\end{equation}
where $c$ is a dimensionless parameter accounting for certain indeterminacies in the effective quantum field theory, $M_{\rm pl}$ is the reduced Planck mass defined by $M_{\rm pl}^2 = (8\pi G)^{-1}$, and $L$ represents the infrared cutoff within the theory. Should $L$ be regarded as the scale of the current universe, such as the Hubble radius $H^{-1}$, the derived dark energy density aligns well with observational data. However, \citet{Hsu:2004ri} noted that this approach yields an incorrect EoS for dark energy. \citet{Li:2004rb} then proposed that $L$ should instead be equated to the scale of the future event horizon,
\begin{equation}\label{2.2}
L=a\int^{\infty}_{t}\frac{{\rm d}t}{a}=a\int^{\infty}_{a}\frac{{\rm d}a}{Ha^{2}},
\end{equation}
with $H(a)$ being the Hubble parameter as a function of the scale factor $a$. This selection not only provides a plausible value for the density of dark energy but also leads to an accelerating universe. Furthermore, the HDE model offers a theoretical explanation for the cosmic coincidence problem \cite{Li:2004rb}.

Investigations into HDE models have been pursued in the literature (see, e.g., Refs.~\cite{Huang:2004wt,Wang:2004nqa,Nojiri:2005pu,Zhang:2005hs,Zhang:2006av,Zhang:2006qu,Setare:2006yj,Zhang:2007sh,Zhang:2007an,Zhang:2014ija,Landim:2015hqa,Wang:2016och,Drepanou:2021jiv,Wang:2023gov}). The parameter $c$ plays a pivotal role in dictating the HDE evolution. At $c = 1$, the EoS of HDE approaches that of a cosmological constant, steering the universe toward a de Sitter phase in the distant future. When $c > 1$, the EoS of HDE stays above $-1$, with HDE exhibiting quintessence-like dark energy behavior. Conversely, if $c < 1$, the EoS of HDE will cross the phantom divide at $w = -1$, culminating in a phantom universe destined for a catastrophic big rip. Prior observational constraints on the HDE model have collectively suggested $c < 1$, hinting that HDE could precipitate a phantom universe ending in a big rip. To circumvent this conundrum, researchers have contemplated the implications of extra dimensions \cite{Zhang:2009xj} or entertained the notion of an interaction between dark energy and dark matter within the HDE model \cite{Li:2008zq}. Moreover, the possibility of direct interactions between dark energy and dark matter might offer a solution or at least a mitigation to the so-called Hubble tension and cosmic coincidence problems. Interacting dark energy (IDE) models have garnered substantial attention \cite{Zhang:2005rj,Zhang:2007uh,Szydlowski:2008by,Zhang:2009qa,Valiviita:2009nu,Li:2010ak,Li:2011ga,Fu:2011ab,Sharif:2012ua,Zhang:2017ize,DiValentino:2017iww,Li:2019ajo,DiValentino:2019ffd,Zhang:2021yof,Wang:2021kxc,Nunes:2022bhn,Zhao:2022bpd,Han:2023exn,Forconi:2023hsj,Li:2023gtu,Halder:2024uao,Benisty:2024lmj,Nong:2024bkr}. The exploration of interacting models within the framework of HDE has been extensively studied (see, e.g., Refs.~\cite{Wang:2005jx,Li:2009zs,Zhang:2012uu,Feng:2016djj,Li:2017usw,Aditya:2019bbk,Sadri:2019qxt}).

Given that the value of $c$ cannot be ascertained from the theoretical framework of the HDE and IHDE models, it is essential to pin down the value of $c$ through cosmological observations. In this study, we harness the latest observational data, encompassing the DESI BAO data, CMB data from Planck and ACT, and SN data from three distinct compilations --- PantheonPlus, Union3, and DESY5 --- to constrain both the HDE and IHDE models. Our primary objective is to ascertain the value of $c$ within the context of the HDE and IHDE models using the latest cosmological observations. Furthermore, we employ Bayesian evidence to assess the extent to which the HDE and IHDE models are endorsed by both early and late-universe data.

This work is organized as follows. In Sec.~\ref{sec2}, we briefly introduce the HDE and IHDE models, as well as the cosmological data used in this work. In Sec.~\ref{sec3}, we report the constraint results and make some relevant discussions. The conclusion is given in Sec.~\ref{sec4}.

\section{methodology and data}\label{sec2}

\subsection{Brief description of the HDE and IHDE models}\label{sec2.1}

The Friedmann equation can be written as
\begin{align}\label{2.1}
3M^2_{\rm{pl}} H^2=\rho_{\rm c}+\rho_{\rm{de}}+... ,
\end{align}
where $3M^2_{\rm{pl}} H^2$ is the critical density of the universe, $\rho_{\rm c}$, $\rho_{\rm{de}}$, and ``...'' represent the energy densities of cold dark matter, dark energy, and rest of the energy densities, respectively.
Then from the energy conservation equations, we have
\begin{align}\label{2.2}
\dot{\rho}_{\rm de} +3H(1+w)\rho_{\rm de}= Q,\\
\dot{\rho}_{\rm c} +3 H \rho_{\rm c}= -Q,
\end{align}
where the dot is the derivative with respect to the cosmic time $t$, $H$ is the Hubble parameter, $w$ is the EoS of dark energy, and $Q$ denotes the phenomenological interaction term describing the energy transfer rate between dark
energy and dark matter due to the interaction.
For the HDE model, $Q = 0$, the EoS of dark energy can be given by
\begin{align}\label{2.3}
w=-\frac{1}{3}-\frac{2\sqrt{\Omega_{\rm{de}}}}{3c} ,
\end{align}
where $\Omega_{\rm{de}} = \rho_{\rm de}/3M^2_{\rm{pl}} H^2$ is the fractional density of dark energy. For the IHDE model, $Q \neq 0$, we have adopted the form $Q=\beta H\rho_{\rm{de}}$, which is currently favored by the observational data \cite{Li:2024qso}, where $\beta$ is a dimensionless coupling parameter describing the strength of interaction between dark energy and dark matter. Note that, according to our convention, $\beta > 0$ means that dark matter decays into dark energy, and $\beta < 0$ means that dark energy decays into dark matter. 

Usually, we face the instability issue in the HDE model when the EoS $w$ crosses the phantom divide $w = -1$, and the IDE model may experience an early-time large-scale instability problem \cite{Majerotto:2009zz,Clemson:2011an}. This instability arises because cosmological perturbations of dark energy within the IDE model diverge in certain regions of the parameter space, potentially leading to the breakdown of IDE cosmology at the perturbation level. To avoid this problem, the parameterized post-Friedmann (PPF) approach \cite{Fang:2008sn,Hu:2008zd} was extended to the IDE models \cite{Li:2014eha,Li:2014cee,Li:2023fdk}, referred to as the ePPF approach. It has been shown that using the ePPF framework can successfully solve the instability issue in IDE cosmology \cite{Zhang:2017ize,Feng:2018yew}. Therefore, we apply the ePPF approach to the HDE and IHDE models to account for the perturbations of dark energy.

\subsection{Cosmological data}\label{sec2.2}
\begin{table}[!htb]
\caption{\label{tab1}Summary of the 3 cosmological models considered in this work. The parameter set for the $\Lambda$CDM model is $\bm{\theta}_{\Lambda\mathrm{CDM}}=\{\Omega_{\rm b} h^2$, $\Omega_{\rm c} h^2$, $\log(10^{10} A_{\mathrm{s}})$, $100\theta_\mathrm{MC}$, $n_{\mathrm{s}}$, $\tau_{\rm reio}\}$}.
\centering
\setlength\tabcolsep{15pt}{
\renewcommand\arraystretch{1.5}
\begin{tabular}{l|cl}
\hline\hline
Model      &   Parameter \texttt{\#} & Free parameters   \\
\hline
$\Lambda$CDM   & 6 & $\bm{\theta}_{\Lambda\mathrm{CDM}}$             \\
HDE  & 7 & $\bm{\theta}_{\Lambda\mathrm{CDM}}$, $c$          \\
IHDE  & 8 & $\bm{\theta}_{\Lambda\mathrm{CDM}}$, $c$, $\beta$     \\

\hline

\end{tabular}}
\end{table}

In this work, we implement the theoretical model in a modified version of the Boltzmann solver CAMB\footnote{\url{https://github.com/liaocrane/IDECAMB/}} \cite{Gelman:1992zz,Li:2023fdk} and use
the publicly available package {\tt Cobaya}\footnote{\url{https://github.com/CobayaSampler/cobaya}} \cite{Torrado:2020dgo} to perform Markov Chain Monte Carlo (MCMC) analyses \cite{Lewis:2002ah,Lewis:2013hha}. We assess the convergence of the MCMC chains using the Gelman-Rubin statistics quantity $R - 1 < 0.02$ \cite{Gelman:1992zz}. The MCMC chains are analyzed using the public package {\tt GetDist}\footnote{\url{https://github.com/cmbant/getdist/}} \cite{Lewis:2019xzd}. In Table~\ref{tab1} we present the basic parameter space for the $\Lambda$CDM, HDE, and IHDE models. We use the current observational data to constrain these models and obtain the best-fit values and the $1$--$2\sigma$ confidence level ranges for the parameters of interest \{$H_{0}$, $\Omega_{\mathrm{m}}$, $c$, $\beta$\}. We adopt the following observational datasets:

$\bullet$ Cosmic Microwave Background (CMB). Measurements of the Planck CMB temperature anisotropy and polarization power spectra, their cross-spectra, and the combination of the ACT and Planck lensing power spectrum. This work employs the following four components of CMB likelihoods: (i) the power spectra of temperature and polarization anisotropies, $C_{\ell}^{TT}$, $C_{\ell}^{TE}$, and $C_{\ell}^{EE}$, at small scales ($\ell > 30$), are obtained from measurements using the Planck \texttt{plik} likelihood~\cite{Planck:2018vyg,Planck:2019nip}; (ii) the spectrum of temperature anisotropies, $C_{\ell}^{TT}$, at large scales ($2 \leq \ell \leq 30$), is obtained from measurements using the Planck \texttt{Commander} likelihood~\cite{Planck:2018vyg,Planck:2019nip}; (iii) the spectrum of E-mode polarization, $C_{\ell}^{EE}$, at large scales ($2 \leq \ell \leq 30$), is obtained from measurements using the Planck \texttt{SimAll} likelihood~\cite{Planck:2018vyg,Planck:2019nip}; (iv) the CMB lensing likelihood, with the latest and most precise data coming from the combination of the NPIPE PR4 Planck CMB lensing reconstruction\footnote{The likelihood is available at \url{https://github.com/carronj/planck_PR4_lensing}.} \citep{Carron:2022eyg} and Data Release 6 of the ACT\footnote{The likelihood is available at \url{https://github.com/ACTCollaboration/act_dr6_lenslike}.} \citep{ACT:2023dou}. We label all the CMB likelihoods as ``CMB".

$\bullet$ Baryon Acoustic Oscillations (BAO). BAO provides a standard ruler for a typical length scale to measure the angular diameter distance $D_{\mathrm{A}}(z)$ and Hubble parameter $H(z)$. The DESI BAO data include tracers of the bright galaxy sample (BGS), luminous red galaxies (LRG), emission line galaxies (ELG), quasars (QSO), and the Lyman-$\alpha$ forest in a redshift range $0.1\leq z \leq 4.2$ \citep{DESI:2024uvr,DESI:2024lzq}. These tracers are described through the transverse comoving distance $D_{\mathrm{M}}/r_{\mathrm{d}}$, the angle-averaged distance $D_{\mathrm{V}}/r_{\mathrm{d}}$, where  $r_{\mathrm{d}}$ is the comoving sound horizon at the drag epoch, and the Hubble horizon $D_{\mathrm{H}}/r_{\mathrm{d}}$. Specifically, we use 12 DESI BAO measurements from Ref.~\cite{DESI:2024mwx}.\footnote{The DESI BAO data used in this work was made public with Data Release 1 (details at \url{https://data.desi.lbl.gov/doc/releases/}).}

\begin{table*}[t]
\caption{Fitting results (68.3\% confidence level) in the $\Lambda$CDM and HDE models from the CMB, DESI, PantheonPlus, Union3, DESY5, CMB+DESI, CMB+PantheonPlus, CMB+Union3, and CMB+DESY5 data. Here, $H_{0}$ is in units of ${\rm km}~{\rm s}^{-1}~{\rm Mpc}^{-1}$.}
\renewcommand\arraystretch{1.5}
\setlength{\tabcolsep}{12pt}
\centering
\resizebox{\textwidth}{!}{
\begin{tabular}{lccccc}
\hline\hline
\multirow{2}{*}{Data} & \multicolumn{2}{c}{$\Lambda$CDM} & \multicolumn{3}{c}{HDE} \\
\cmidrule[0.5pt](l{2pt}r{2pt}){2-3} \cmidrule[0.5pt](l{2pt}r{2pt}){4-6}
& $ H_0 $ & $ \Omega_\mathrm{m} $ & $ H_0 $ & $ \Omega_\mathrm{m} $ & $ c $  \\ 
\hline
CMB    & $67.21\pm 0.45$ & $0.316\pm 0.006$ & $77.50^{+7.90}_{-4.00}$ & $0.243^{+0.019}_{-0.051}$ & $ 0.452^{+0.023}_{-0.093}$  \\
DESI   & --- & $0.295\pm 0.015$ & --- & $0.271^{+0.014}_{-0.017}$ & $0.890^{+0.130}_{-0.250}$ \\
PantheonPlus   & --- & $0.332\pm 0.018$ & --- & $0.222^{+0.047}_{-0.061}$ & $1.330^{+0.300}_{-0.460}$  \\
Union3 & --- & $0.358\pm 0.026$ & --- & $0.197^{+0.037}_{-0.096}$ & $1.870^{+0.570}_{-0.650}$  \\
DESY5 & --- & $0.352\pm 0.017$ & --- & $0.201^{+0.031}_{-0.089}$ & $1.630^{+0.420}_{-0.570}$  \\
CMB+DESI & $67.86\pm 0.37$ & $0.307\pm 0.005$ & $76.80\pm 2.30$  & $0.241^{+0.013}_{-0.016}$ & $0.462^{+0.027}_{-0.036}$  \\
CMB+PantheonPlus & $67.12\pm 0.44$ & $0.317\pm 0.059$ & $64.43\pm 0.75$ & $0.346\pm 0.009$ & $0.688^{+0.026}_{-0.030}$  \\
CMB+Union3 & $67.04\pm 0.45$ & $0.318\pm 0.006$ & $63.80\pm 0.97$ & $0.354\pm 0.012$ & $0.713^{+0.039}_{-0.045}$  \\
CMB+DESY5 & $66.88\pm 0.42$ & $0.320\pm 0.058$ & $63.76\pm 0.68$ & $0.354\pm 0.008$  & $0.711^{+0.026}_{-0.030}$   \\
\hline
\end{tabular}
}
\label{tab2}              
\end{table*}

\begin{figure*}[t]
\resizebox{\textwidth}{!}{
\begin{minipage}{0.48\textwidth}
\centering
\includegraphics[width=\linewidth]{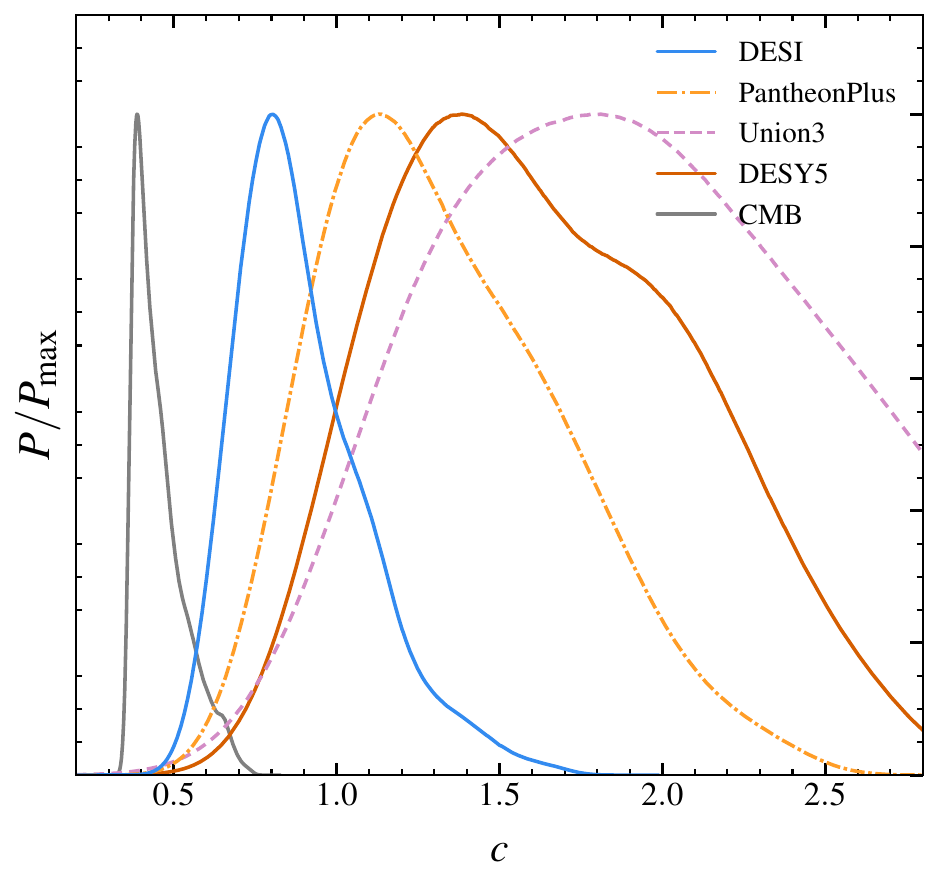}
\end{minipage}\hfill
\begin{minipage}{0.46\textwidth}
\centering
\includegraphics[width=\linewidth]{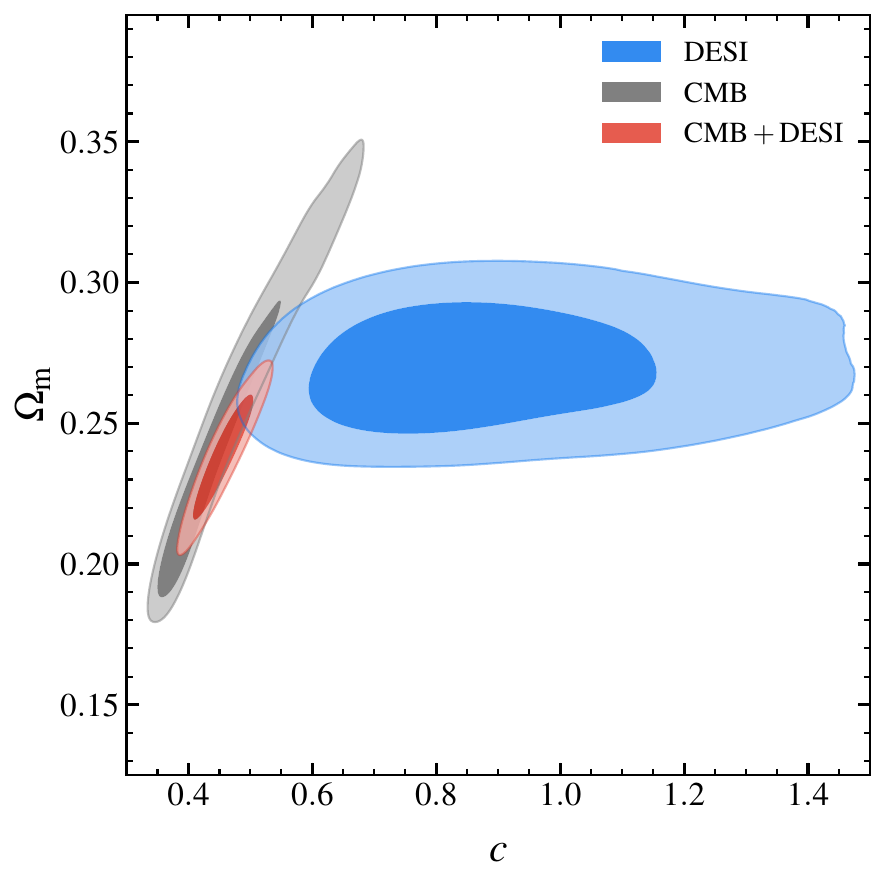}
\end{minipage}
\begin{minipage}{0.468\textwidth}
\centering
\includegraphics[width=\linewidth]{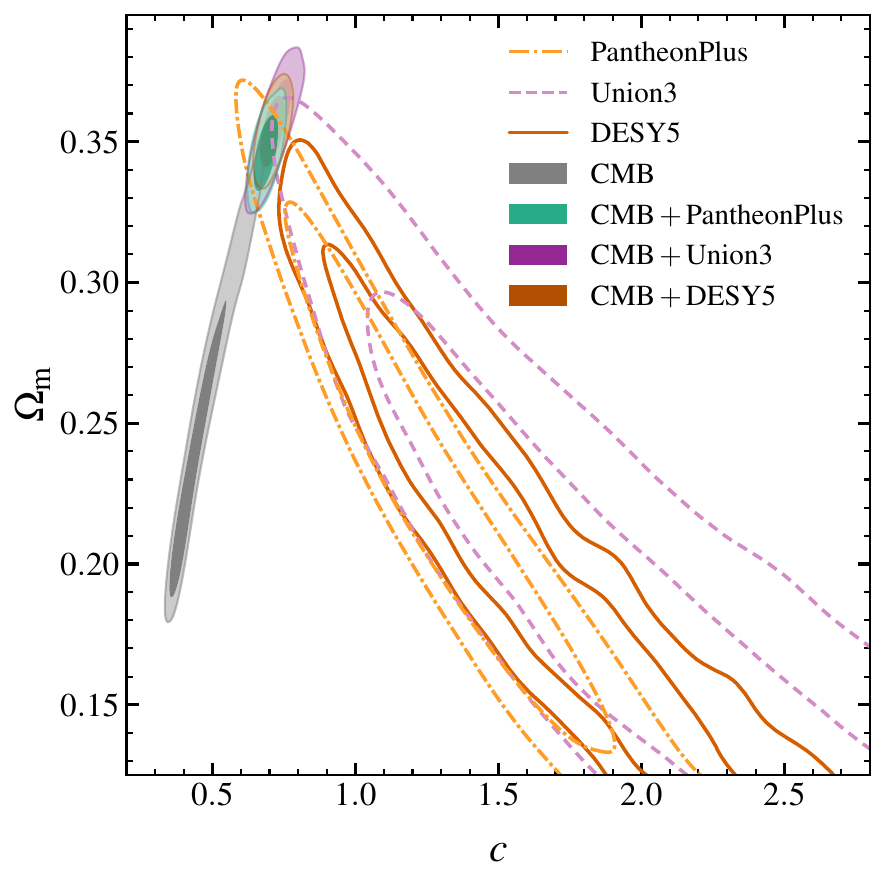}
\end{minipage}
}
\centering \caption{\label{fig1} Constraints on cosmological parameters using CMB, BAO, and SN data. Left panel: The one-dimensional (1D) marginalized posterior constraints on $c$ using DESI, PantheonPlus, Union3, DESY5, and CMB data in HDE model. Middle panel: Two-dimensional marginalized contours (68.3\% and 95.4\% confidence level) in the $c$--$\Omega_{\rm m}$ plane by using the DESI, CMB, and CMB+DESI data in the HDE model. Right panel: Two-dimensional marginalized contours (68.3\% and 95.4\% confidence level) in the $c$--$\Omega_{\rm m}$ plane by using the PantheonPlus, Union3, DESY5, CMB, CMB+PantheonPus, CMB+Union3, and CMB+DESY5 data in the HDE model.}
\end{figure*}

$\bullet$ Type Ia Supernovae (SNe). 
SNe can effectively probe the late-time cosmic evolution through the relation $\mu(z) = 5\log_{10} d_\mathrm{L}(z) + \text{cst}$, where $d_\mathrm{L}(z) \equiv H_0 D_\mathrm{L}(z)/c$ is the dimensionless luminosity distance, $\text{cst}$ is a constant (representing the degeneracy between $H_0$ and the SNe absolute magnitude $M_B$). Here, $D_{\mathrm{L}}(z) = (1+z) D_{\mathrm{M}}(z)$ is the luminosity distance. We adopt SN data from three compilations: (i) the PantheonPlus comprises 1550 spectroscopically confirmed SNe from 18 different surveys, spanning the redshift range of $0.01 < z < 2.26$ \cite{Brout:2022vxf}; (ii) the Dark Energy Survey, as part of their Year 5 data release, recently published results based on a new, homogeneously selected sample includes 194 low-redshift SNe ($0.025 < z < 0.1$) and 1635 SNe classified photometrically, covering the range $0.1 < z < 1.3$ \cite{DES:2024tys}; (iii) the more recent Union3 compilation, presented in Ref.~\cite{Rubin:2023ovl}, includes 2087 SNe, with many (1363 SNe) overlapping with those in PantheonPlus.

\section{Results and discussions}\label{sec3}

\begin{figure*}[!htp]
\includegraphics[width=0.45\textwidth]{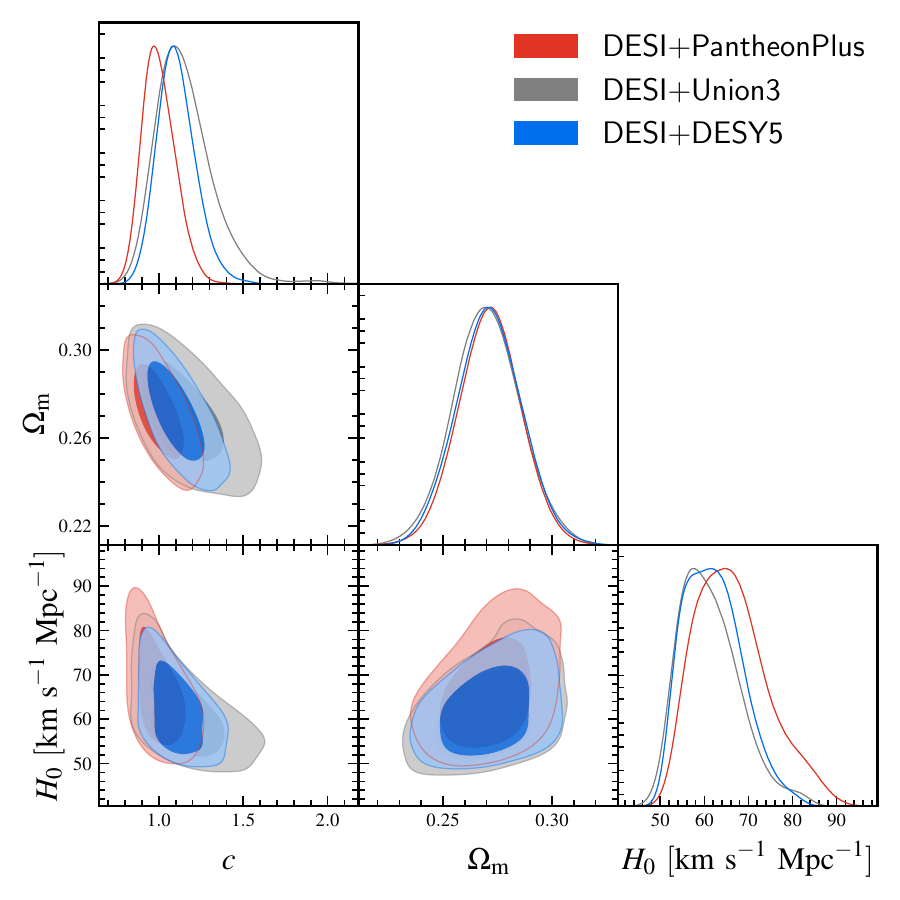} \ \hspace{1cm}
\includegraphics[width=0.45\textwidth]{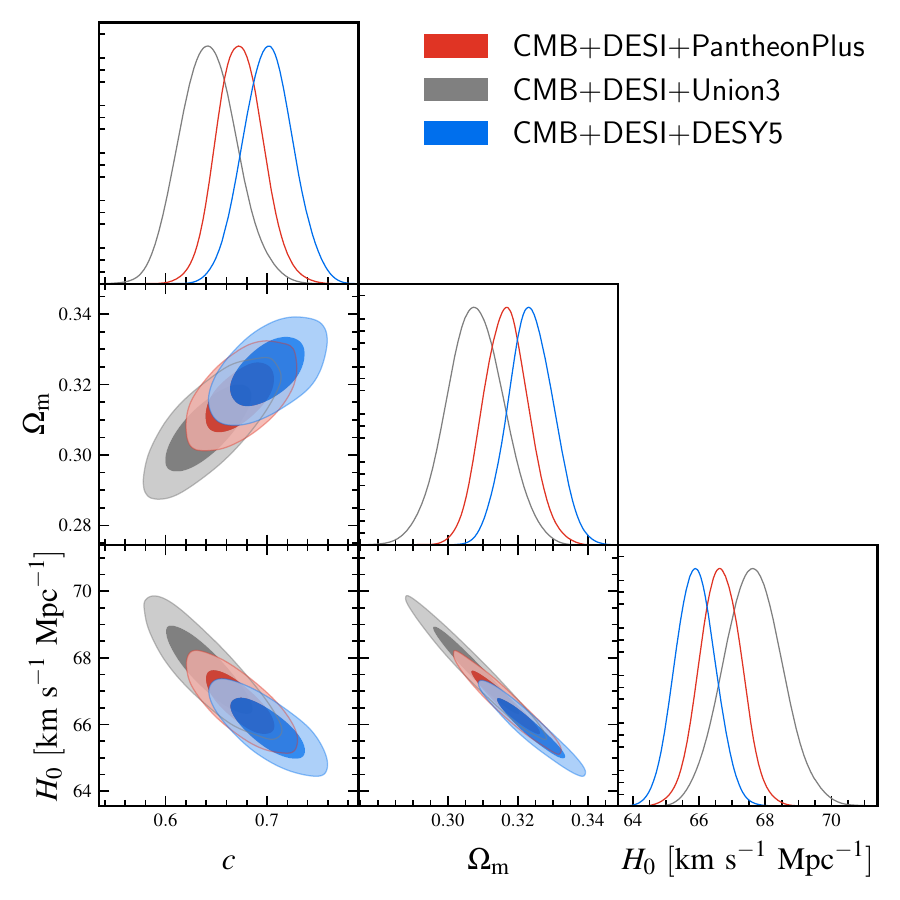}\ \hspace{1cm}
\centering \caption{\label{fig2} {Constraints on cosmological parameters using CMB, BAO, and SN data. Left panel: Constraints on the cosmological parameters using the DESI+PantheonPlus, DESI+Union3, and DESI+DESY5 data in the HDE model. Right panel: Constraints on the cosmological parameters using the CMB+DESI+PantheonPlus, CMB+DESI+Union3, and CMB+DESI+DESY5 data in the HDE model.}}
\end{figure*}

\begin{figure*}[!htp]
\includegraphics[width=0.45\textwidth]{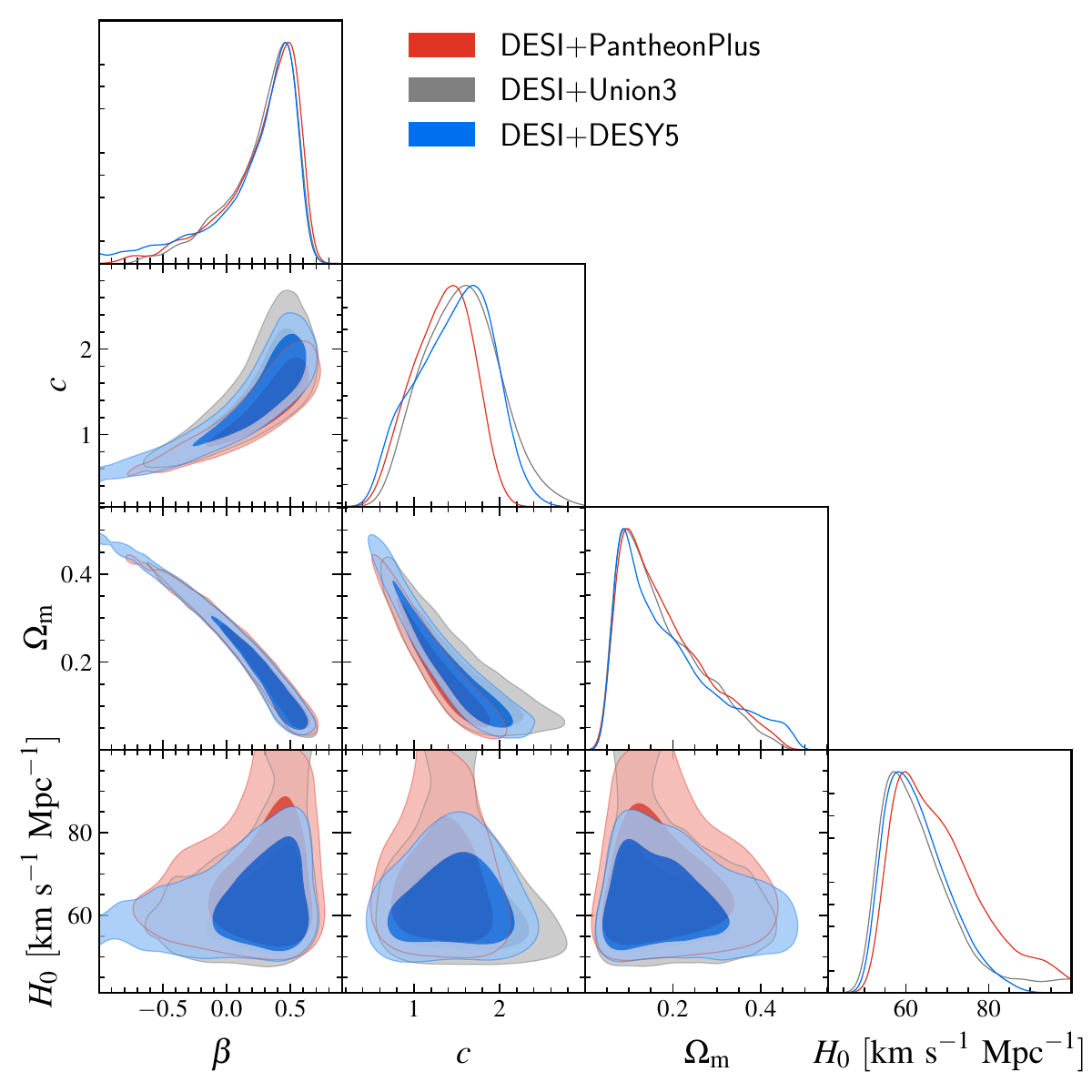} \ \hspace{1cm}
\includegraphics[width=0.45\textwidth]{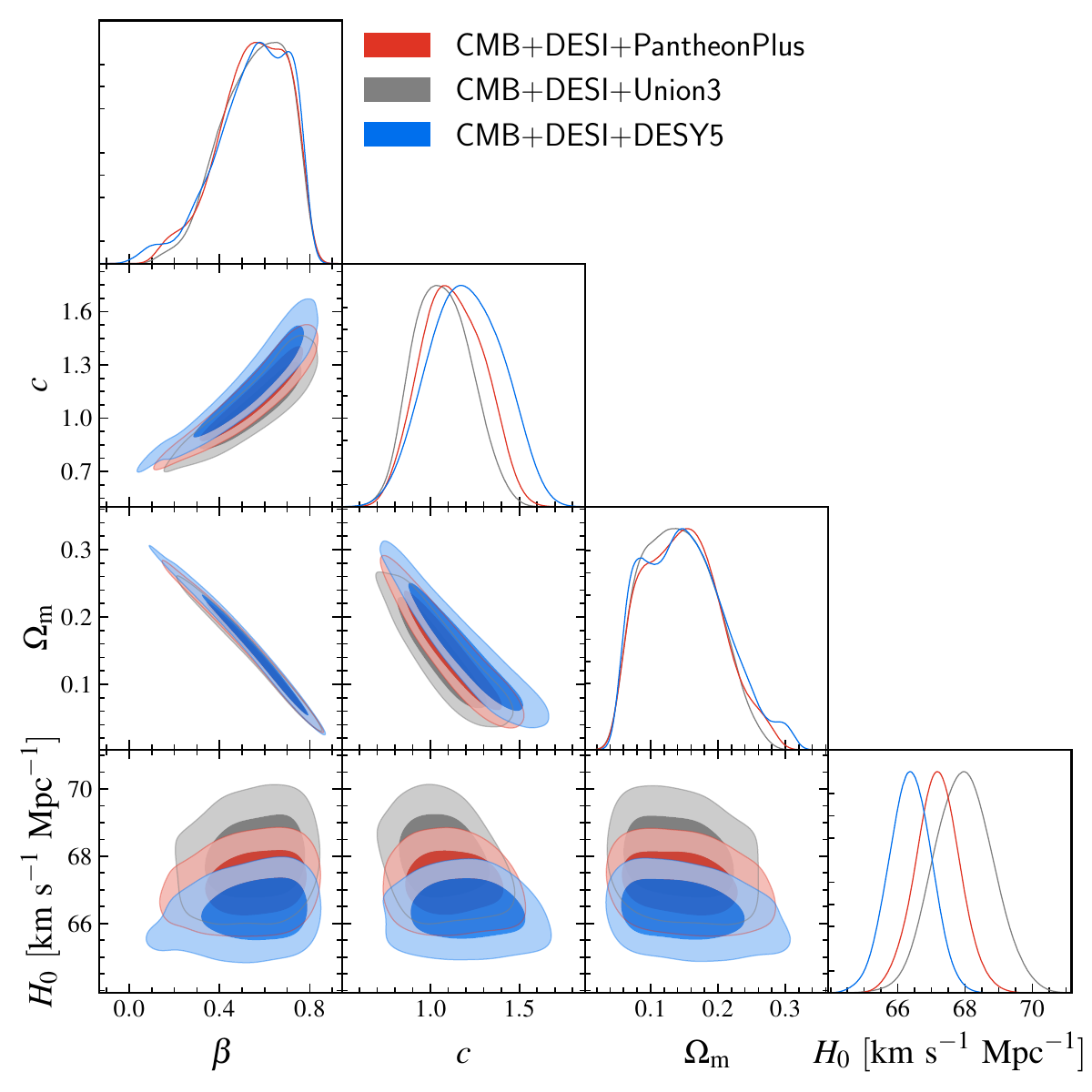}\ \hspace{1cm}
\centering \caption{\label{fig3} {Constraints on cosmological parameters using CMB, BAO, and SN data. Left panel: Constraints on the cosmological parameters using the DESI+PantheonPlus, DESI+Union3, and DESI+DESY5 data in the IHDE model. Right panel: Constraints on the cosmological parameters using the CMB+DESI+PantheonPlus, CMB+DESI+Union3, and CMB+DESI+DESY5 data in the IHDE model.}}
\end{figure*}

\begin{figure*}[!htp]
\includegraphics[width=\textwidth]{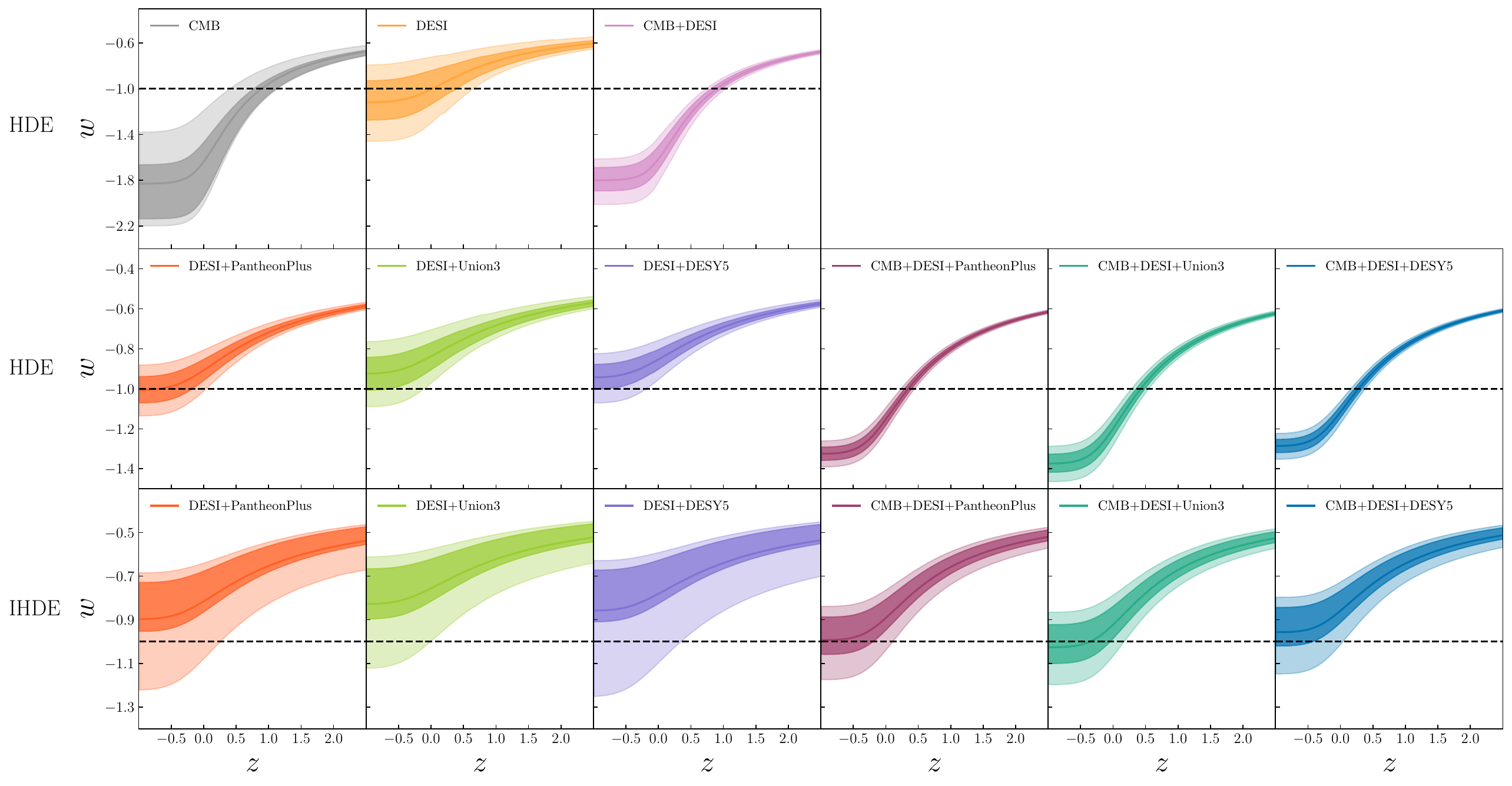}
\centering
\caption{\label{fig4} Reconstructed evolution history of $w$ at $1\sigma$ and $2\sigma$ confidence levels in the HDE and IHDE models, constrained by current observational data. The black dashed line denotes the cosmological constant boundary $w=-1$.}
\end{figure*}

\begin{figure*}[!htp]
\includegraphics[width=\textwidth]{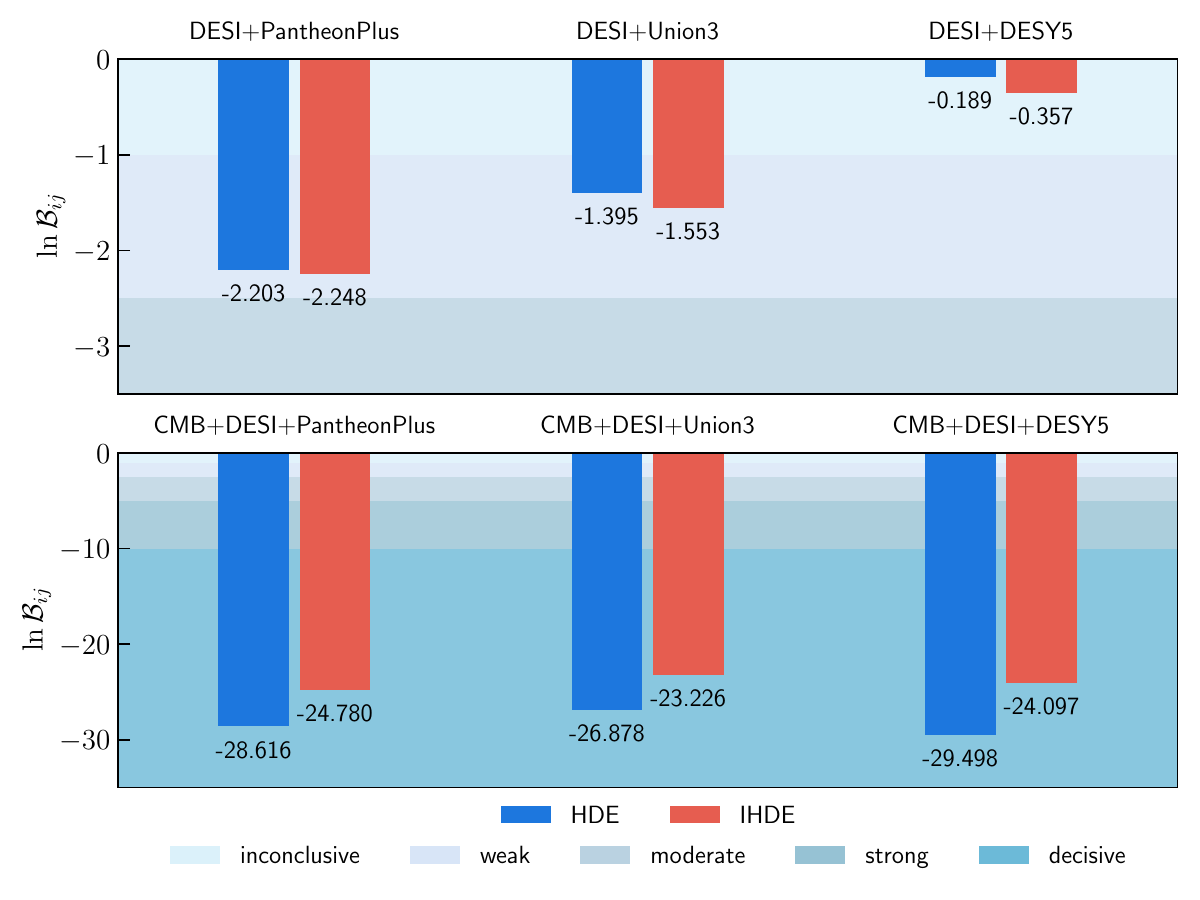}
\centering
\caption{\label{fig5} Comparison of the Bayesian evidence for the HDE and IHDE models and the $\Lambda$CDM model using current observational data. {The Bayes factor $\ln \mathcal{B}_{ij}$ (where $i$ = HDE or IHDE, $j$ = $\Lambda$CDM) and its strength according to Jeffreys' scale are used to assess the preference between models, where a negative value indicates a preference for the $\Lambda$CDM model.}}
\end{figure*}

In this section, we shall report the constraint results of the cosmological parameters. We consider the HDE and IHDE models to perform a cosmological analysis using current observational data, including DESI, CMB, PantheonPlus, Union3, and DESY5 data. We show the $1\sigma$ and $2\sigma$ posterior distribution contours for various cosmological parameters in the HDE and IHDE models, as shown in Figs.~\ref{fig1}--\ref{fig3}. The $1\sigma$ errors for the marginalized parameter constraints are summarized in Tables~\ref{tab2} and \ref{tab3}. We reconstructed the evolution history of $w$ at the $1\sigma$ and $2\sigma$ confidence levels for the HDE and IHDE models, constrained by the current observational data, as shown in Fig.~\ref{fig4}. Finally, we compared $\ln \mathcal{B}_{ij}$ between the models using the current observational data, as shown in Fig.~\ref{fig5}.

In the left panel of Fig.~\ref{fig1}, we present the constraint results of the DESI, PantheonPlus, Union3, DESY5, and CMB data in the 1D marginalized posterior constraints on $c$, and we also provide the constraint results in Table~\ref{tab2}. It is evident that the values of the parameter $c$ derived from the three datasets are different, with the BAO and SN datasets yielding particularly higher values compared to the CMB data. In this work, we primarily focus on the constraints on the parameter $c$ provided by the combination of datasets. Therefore, quantifying the level of consistency between the different datasets is of significant importance to ensure the safe combination of the data. We do so by calculating the tension between the two datasets as \cite{Camera:2017tws,DES:2020hen}
\begin{align}\label{2.4}
T_c = \frac{|c_{\text{m}} - c_{\text{n}}|}{\sqrt{\sigma_{\text{m}}^2 + \sigma_{\text{n}}^2}},
\end{align}
where $\text{m}$ and $\text{n}$ represent different datasets. It is worth noting that the constraint results for the parameter $c$ are given by $c^{+\sigma_{+}}_{-\sigma_{-}}$, which reflects the asymmetric uncertainties. For the case of $c_{\text{m}} > c_{\text{n}}$, we use $\sigma_{\text{m}+}$ and $\sigma_{\text{n}-}$ in the calculation of $T_c$.

Applying this statistic to DESI with PantheonPlus, Union3, and DESY5, we find that the differences in the parameter $c$ correspond to mild discrepancies of $1.13\sigma$, $1.57\sigma$, and $1.51\sigma$, respectively. Therefore, we conclude that the DESI BAO data are consistent with the PantheonPlus, Union3, and DESY5 datasets in this parameter space. The statistical significance of the differences with the CMB, calculated as described above, is $2.74\sigma$ for DESI, $2.79\sigma$ for PantheonPlus, $2.45\sigma$ for Union3, and $2.74\sigma$ for DESY5. Although these values indicate a degree of disagreement between these datasets and the CMB results, they do not meet the $3\sigma$ threshold for significant tension. Therefore, we believe it is safe to combine these datasets.

\begin{table*}[!htb]
\renewcommand\arraystretch{2}
\centering
\scriptsize
\caption{Fitting results (68.3\% confidence level) in the $\Lambda$CDM, HDE, and IHDE models from the DESI+PantheonPlus, DESI+Union3, DESI+DESY5, CMB+DESI+PantheonPlus, CMB+DESI+Union3, and CMB+DESI+DESY5 data. Here, $H_{0}$ is in units of ${\rm km}~{\rm s}^{-1}~{\rm Mpc}^{-1}$.}
\label{tab3}
\begin{tabular}{lccccccc} 
\hline
\hline
Model&Parameter & DESI+PantheonPlus & DESI+Union3 & DESI+DESY5 & CMB+DESI+PantheonPlus & CMB+DESI+Union3 & CMB+DESI+DESY5 \\
\hline
$\Lambda$CDM& $H_{0}$ & $-$ & $-$ & $-$  & $67.73\pm 0.36$ & $67.72\pm 0.35$ & $67.57\pm 0.34$ \\
 & $\Omega _{\rm m}$ & $0.311\pm 0.012$ & $0.316\pm 0.013$ & $0.322\pm 0.011$  & $0.309\pm 0.005$ & $0.309\pm 0.005$ & $0.310\pm 0.005$ \\
HDE	& $H_{0}$ & $66.20^{+5.90}_{-10.00}$ & $61.20^{+5.00}_{-8.20}$ & $62.00^{+5.20}_{-8.10}$  & $66.65\pm 0.63$ & $67.64\pm 0.88$ & $65.89\pm 0.60$ \\
& $\Omega _{\rm m}$ & $0.271\pm 0.014$ & $0.270\pm 0.015$ & $0.271\pm 0.015$  & $0.317\pm 0.006$ & $0.308\pm 0.008$ & $0.324\pm 0.006$ \\
& $c$ & $1.001^{+0.083}_{-0.110}$ & $ 1.150^{+0.121}_{-0.190}$ & $ 1.106^{+0.096}_{-0.130}$  & $0.673\pm 0.023$ & $0.642\pm 0.028$ & $0.701\pm 0.024$ \\
IHDE & $H_{0}$ & $68.20^{+5.90}_{-13.00}$ & $63.50^{+4.30}_{-11.00}$ & $63.00^{+4.90}_{-9.50}$  & $67.18\pm 0.67$ & $67.95\pm 0.87$ & $66.36\pm 0.63$ \\
& $\Omega _{\rm m}$ & $0.178^{+0.047}_{-0.120}$ & $0.173^{+0.047}_{-0.121}$ & $0.186^{+0.045}_{-0.130}$  & $0.147^{+0.047}_{-0.068}$ & $0.142^{+0.046}_{-0.065}$ & $0.150^{+0.042}_{-0.081}$ \\
& $c$ & $1.33^{+0.40}_{-0.32}$ & $ 1.55^{+0.44}_{-0.52}$ & $ 1.47^{+0.53}_{-0.38}$  & $1.12\pm 0.17$ & $1.07\pm 0.16$ & $1.20\pm 0.20$ \\
& $\beta$ & $0.27^{+0.35}_{-0.11}$ & $0.27^{+0.33}_{-0.11}$ & $ 0.26^{+0.38}_{-0.09}$  & $ 0.55^{+0.19}_{-0.11}$ & $ 0.56^{+0.18}_{-0.11}$ & $ 0.55^{+0.21}_{-0.10}$ \\	
\hline
\end{tabular}
\end{table*}

In the middle panel of Fig.~\ref{fig1}, we show the constraint results of the DESI, CMB, and CMB+DESI data in the $c$--$\Omega_{\rm m}$ plane for the HDE model. The constraint values of $c$ are  $0.890^{+0.130}_{-0.250}$, $0.452^{+0.023}_{-0.093}$, and $0.462^{+0.027}_{-0.036}$ for DESI, CMB, and CMB+DESI data, respectively. We find that the best-fit value of $c$ obtained from DESI data alone is relatively large, while the value derived from CMB data alone is smaller, with only a slight improvement observed after adding DESI data. The combination of DESI data and CMB data can effectively break the cosmological parameter degeneracies. Our results show that in the HDE model, CMB+DESI data result in $\sigma(\Omega_{\rm m})=0.0145$ and $\sigma($c$)=0.0315$, which are 58.5\% and 45.6\% better than those of CMB data. Moreover, we find that in the HDE model, the CMB and CMB+DESI data yield smaller values for $\Omega_{\rm m}$ and larger values for $H_0$ compared to the results in the $\Lambda$CDM model. In the right panel of Fig.~\ref{fig1}, we show the constraint results of the SN, CMB, and CMB+SN data in the $c$--$\Omega_{\rm m}$ plane for the HDE model. We find that the best-fit value of $c$ obtained from SN data alone is relatively large, and although it decreases when the CMB data is included, it still remains higher than the result from the CMB+DESI data. The parameter constraint precision achieved with the CMB+SN and CMB+DESI data is comparable in terms of improvement relative to the CMB alone. Interestingly, we find that in the HDE model, the CMB+SN data yield larger values for $\Omega_{\rm m}$ and smaller values for $H_0$ compared to the results in the $\Lambda$CDM model, which contrasts with the CMB+DESI results.

In the left panel of Fig.~\ref{fig2}, we show the constraint results for the HDE model based on DESI BAO combined with SN data. The constraint values of $c$ are $1.001^{+0.083}_{-0.110}$, $1.150^{+0.121}_{-0.190}$, and $1.106^{+0.096}_{-0.130}$ for DESI+PantheonPlus, DESI+Union3, and DESI+DESY5, respectively. We find that the combinations of DESI BAO with different SN datasets exhibit moderate variations in both central values and quoted uncertainties, but all of them result in $c > 1$ within approximately the $1\sigma$ level. In the right panel of Fig.~\ref{fig2}, we show the constraint results of the combinations of CMB with DESI+PantheonPlus, DESI+Union3, and DESI+DESY5, respectively. The constraint values of $c$ are $0.673 \pm 0.023$, $0.642 \pm 0.028$, and $0.701 \pm 0.024$ for CMB+DESI+PantheonPlus, CMB+DESI+Union3, and CMB+DESI+DESY5, respectively. We find that adding CMB data significantly decreases the value of $c$, resulting in $c < 1$ beyond the $10\sigma$ level. This indicates that in the HDE model, CMB data have a strong impact on the parameter $c$, whereas the combined results of CMB and BAO+SN are less influenced by the SN data.

In the left panel of Fig.~\ref{fig3}, we show the constraint results for the IHDE model based on DESI BAO combined with SN data. The constraint values of $c$ are $1.33^{+0.40}_{-0.32}$, $1.55^{+0.44}_{-0.52}$, and $1.47^{+0.53}_{-0.38}$, using DESI+PantheonPlus, DESI+Union3, and DESI+DESY5, respectively. We find that the value of $c$ in the IHDE model is significantly higher than in the HDE model, and $c > 1$ can be obtained at the $1\sigma$ level. In the right panel of Fig.~\ref{fig3}, we show the constraint results of the combination of CMB with DESI+PantheonPlus, DESI+Union3, and DESI+DESY5, respectively. The constraint values of $c$ are $1.12 \pm 0.17$, $1.07 \pm 0.16$, and $1.20 \pm 0.20$, using CMB+DESI+PantheonPlus, CMB+DESI+Union3, and CMB+DESI+DESY5, respectively. We find that the results and the associated uncertainties again are weakly influenced by the choice of SN dataset, and the CMB combined with DESI BAO and SN data show a preference for $c > 1$ in the IHDE model. The constraint values of $\beta$ are $0.55^{+0.19}_{-0.11}$, $0.56^{+0.18}_{-0.11}$, and $0.55^{+0.21}_{-0.10}$, using CMB+DESI+PantheonPlus, CMB+DESI+Union3, and CMB+DESI+DESY5, respectively. The results obtained here indicate a positive coupling can be detected with more than $3\sigma$ significance based on the current observational data.

In order to better understand the evolution of dark energy in the HDE and IHDE models, we use current observational data to reconstruct the evolution of $w$ with respect to redshift $z$, as shown in the panels of Fig.~\ref{fig4}. We find that in the HDE model with CMB and CMB+DESI constraints, $w$ crosses $-1$ at more than $2\sigma$ confidence level. As mentioned above, this implies that the future universe will be dominated by phantom energy, leading to the occurrence of the big rip. The situation is slightly better when using DESI alone. When using the combinations of DESI with PantheonPlus, Union3, and DESY5 for the HDE model, $w$ does not cross $-1$ at approximately the $1\sigma$ confidence level. However, once CMB data is included, $w$ crosses $-1$ at a significance level exceeding $10\sigma$, indicating that the big rip will occur. After introducing interaction into the HDE model, using the combinations of DESI with PantheonPlus, Union3, and DESY5 for the IIHDE model, $w$ will not cross $-1$ at the $1\sigma$ confidence level. Even with the addition of CMB data, the risk of $w$ crossing $-1$ is significantly reduced. Therefore, the inclusion of interaction in the HDE model may help resolve or mitigate the cosmic big rip conundrum.

Finally, we employ the Bayesian Evidence selection criterion as a method for selecting the best model. Here, we use publicly available code
{\tt MCEvidence}\footnote{\url{https://github.com/yabebalFantaye/MCEvidence}} \cite{Heavens:2017hkr,Heavens:2017afc} to compute the Bayes factor of the models. The Bayesian evidence $Z$ is given by
\begin{equation}
Z = \int_{\Omega} P(D|\bm{\theta},M)P(\bm{\theta}|M)P(M)\ {\rm d}\bm{\theta},
\label{eq: lnZ}
\end{equation}
where $P(D|\bm{\theta},M)$ is the likelihood of the data $D$ given the parameters $\bm{\theta}$ and the model $M$, $P(\bm{\theta}|M)$ is the prior probability of $\bm{\theta}$ given $M$, and $P(M)$ is the prior of $M$. Then we calculate the Bayes factor $\ln \mathcal{B}_{ij} = \ln Z_i - \ln Z_j$ in logarithmic space, where $Z_i$ and $Z_j$ are Bayesian evidence of two models.

Typically, the Jeffreys scale \cite{Kass:1995loi,Trotta:2008qt} is employed to gauge the strength of model preference: if $\left|\ln \mathcal{B}_{ij}\right|<1$, the evidence is inconclusive; $1\le\left|\ln \mathcal{B}_{ij}\right|<2.5$ represents weak evidence; $2.5\le\left|\ln \mathcal{B}_{ij}\right|<5$ is moderate; $5\le\left|\ln \mathcal{B}_{ij}\right|<10$ is strong; and if $\left|\ln \mathcal{B}_{ij}\right|\ge 10$, the evidence is decisive. It is worth noting that a positive value of $\ln \mathcal{B}_{ij}$ represents a preference for model $i$ over model $j$.

In Fig.~\ref{fig5}, we show the Bayes factors $\ln \mathcal{B}_{ij}$ for the HDE and IHDE models relative to the $\Lambda$CDM model, based on the current observational data. {Here, $i$ denotes the HDE or IHDE model and $j$ denotes the $\Lambda$CDM model}. It is worth emphasizing that negative values indicate a preference for the $\Lambda$CDM model. The combinations of DESI with PantheonPlus, Union3, and DESY5 yield Bayes factors of $|\ln \mathcal{B}_{ij}|$ = 2.203, 1.395, and 0.189 for HDE, and $|\ln \mathcal{B}_{ij}|$ = 2.248, 1.553, and 0.357 for IHDE, relative to $\Lambda$CDM, respectively. When CMB data are added, the combinations CMB+DESI+PantheonPlus, CMB+DESI+Union3, and CMB+DESI+DESY5 yield Bayes factors of $|\ln \mathcal{B}_{ij}|$ = 28.616, 26.878, and 29.498 for HDE, and $|\ln \mathcal{B}_{ij}|$ = 24.780, 23.226, and 24.097 for IHDE, relative to $\Lambda$CDM, respectively. We find that using the combinations of DESI with PantheonPlus, Union3, and DESY5 data shows essentially equal preferences among the models, with $\Lambda$CDM only slightly favored over the extended models. When CMB data are included, the current data combination strongly prefers $\Lambda$CDM over the HDE and IHDE models. In addition, in this case, the HDE model is more strongly disfavored than the IHDE model.

\section{Conclusion}\label{sec4}

The parameter $c$ is a crucial factor that determines the properties of HDE. It is important to note that the value of $c$ cannot be derived from the theoretical framework of the HDE model; instead, it must be obtained by fitting observational data. In this work, we investigate both the HDE and IHDE models using early-universe data from the CMB and the latest late-universe data, specifically SN data from PantheonPlus, Union3, and DESY5, as well as BAO data from DESI 2024. We also aim to perform model selection using Bayesian evidence to evaluate whether the HDE and IHDE models are better supported by the current observational data.

Our findings indicate that the value of $c$ obtained using DESI data alone is higher than that derived from CMB data in the HDE model. When combining CMB and DESI data, the parameter degeneracy can be effectively broken, resulting in approximately a 50\% improvement in parameter constraint capability. We observe that when combining BAO and SN data in the HDE model, $c > 1$ is obtained at approximately the $1\sigma$ level. However, once CMB data are included, $c < 1$ is established beyond the $10\sigma$ level. For the IHDE model, whether combining only BAO and SN data or including CMB data, the parameter $c > 1$ is typically obtained at the $1\sigma$ level in most cases. Additionally, using the combination of BAO and SN data, we find that $w$ does not cross $-1$ within approximately the $1\sigma$ range, indicating that a big rip scenario is unlikely. When CMB data are included, $w$ crosses $-1$ in the HDE model, while in the IHDE model, the risk of $w$ crossing $-1$ is significantly reduced or avoided. Furthermore, current observational data suggest that a positive coupling can be detected with more than $3\sigma$ significance.

We utilize current observational data, including DESI 2024, CMB, PantheonPlus, Union3, and DESY5, to calculate the Bayesian evidence for the HDE and IHDE models. Our analysis reveals that combinations of DESI with PantheonPlus, Union3, and DESY5 data show essentially equal preferences among the models, with $\Lambda$CDM being only slightly favored over the extended models. For instance, the combination of DESI with DESY5 data yields a Bayes factor for HDE relative to 
$\Lambda$CDM of $|\ln \mathcal{B}_{ij}|$ = 0.189. When CMB data are included, the current data combination strongly favors 
$\Lambda$CDM over the HDE and IHDE models. Our results indicate that the HDE and IHDE models merit further exploration using more precise late-universe observations in the future. In the coming years, the complete DESI dataset, along with more accurate late-universe data from EUCLID \cite{Euclid:2024yrr} and LSST \cite{LSST:2008ijt}, will enhance our understanding of the nature of dark energy, thereby advancing the study of HDE.

\section*{Acknowledgments}
We thank Shang-Jie Jin, Hai-Li Li, and Sheng-Han Zhou for their helpful discussions. This work was supported by the National SKA Program of China (Grants Nos. 2022SKA0110200 and 2022SKA0110203), the National Natural Science Foundation of China (Grants Nos. 12473001, 11975072, 11875102, 11835009, and 12305069), the National 111 Project (Grant No. B16009), and the Program of the Education Department of Liaoning Province (Grant No. JYTMS20231695). 

\textbf{Data Availability Statement} This manuscript has no associated data. [Author’s comment: All data analysed in this
 work is publicly available and references have been provided in Sec.~\ref{sec2.2}.]

\textbf{Code Availability Statement} This manuscript has no associated code/software. [Author’s comment: In this work, we utilized the publicly available software packages Cobaya, GetDist, MCEvidence, and the CAMB code. The relevant references are provided in Sec.~\ref{sec2.2} and Sec.~\ref{sec3}. No other code/software was generated or analysed during
this work.]

\bibliography{DESI_HDE}

\end{document}